\documentclass[showpacs,preprint,aps]{revtex4}

\def\Frac#1#2{\frac{\displaystyle{#1}}{\displaystyle{#2}}}

\input epsf

\begin{document}

\preprint{CERN-TH/2002-149 \cr  FNAL-Pub-02/137-A \cr LAPTH-921-02 }

\vspace*{24pt}

\title{Transdimensional physics and inflation}

\author{Gian F.\ Giudice}
\author{Edward W.\ Kolb}
    \altaffiliation[Permanent address: ]{Fermilab Astrophysics Center, Fermi
       National Accelerator Laboratory, Batavia, Illinois \ 60510-0500, USA,
       and Department of Astronomy and Astrophysics, Enrico Fermi Institute,
       The University of Chicago, Chicago, Illinois \ 60637-1433, USA}
\author{Julien Lesgourgues}
   \altaffiliation[Permanent address: ]{LAPTH, Chemin de Bellevue, B.P. 110, 
       F-74941 Annecy-Le-Vieux Cedex, France}
   \affiliation{Theory Division, CERN, CH-1211 Geneva 23, Switzerland}
\author{Antonio Riotto}
   \affiliation{INFN, Sezione di Padova, Via Marzolo 8, Padova I-35131, Italy}

\date{July 2002}
             
\begin{abstract}
Within the framework of a five-dimensional brane world with a stabilized
radion, we compute the cosmological perturbations generated during inflation
and show that the perturbations are a powerful tool to probe the physics of
extra dimensions.  While we find that the power spectrum of scalar
perturbations is unchanged, we show that the existence of the fifth dimension
is imprinted on the spectrum of gravitational waves generated during inflation.
In particular, we find that the tensor perturbations receive a correction
proportional to $(HR)^2$, where $H$ is the Hubble expansion rate during
inflation and $R$ is the size of the extra dimension. We also generalize our
findings to the case of several extra dimensions as well as to warped
geometries.
\end{abstract}


\maketitle

\section{Introduction}

The expanding Universe, especially if it underwent a primordial inflationary
phase \cite{review}, represents the most powerful probe of small distance
scales at our disposal.  Present-day astronomical length scales were extremely
tiny at early epochs and were sensitive to short-distance physics.  This simple
observation has recently generated a lot of excitement about the possibility of
opening a window on transplanckian or stringy physics in Cosmic Microwave
Background (CMB) anisotropies \cite{tp}.  Unfortunately, in the absence of a
quantum theory of gravity, uncontrollable nonlinear effects may dominate at
transplanckian distances, and the behavior of the cosmological perturbations
and crucial related issues such as the definition of the vacuum remain
unknown. This makes it difficult to predict on firm grounds the signatures of
transplanckian physics on present-day cosmological scales \cite{tpf}.

In this paper we will demonstrate that cosmological perturbations generated
during inflation may nevertheless provide a powerful probe of another important
aspect of many modern theories of particle physics: the existence of extra
dimensions.  The presence of extra dimensions is a crucial ingredient in
theories explaining the unification of gravity and gauge forces.  A typical
example is string theory, where more than three spatial dimensions are
necessary for the consistency of the theory. It has recently become clear that
extra dimensions may be very large and could even be testable in accelerator
experiments.

In theories with $n$ compactified extra dimensions with typical radii $R$, the
four-dimensional Planck mass, $M_{P}$, is just a derived quantity, while the
fundamental scale is the gravitational mass, $M_*$, of the $(n+4)$-dimensional
theory. The mass scale $M_*$ is a free parameter and can range from a TeV to
$M_{P}$, with $M_{P}^2\sim M_*^{n+2}\, R^n$. The size of extra dimensions can
range from macroscopic scales down to Planckian distances.

In general there is a large hierarchy between the size of extra dimensions,
$R$, and $M_*^{-1}$, with $R\gg M_*^{-1}$. This means that perturbations that
are currently observable on cosmological scales might have been generated at
early times on scales much smaller than the size of extra dimensions, but still
on scales larger than the fundamental Planck mass so that the
$(4+n)$-dimensional Einstein equations should describe gravity and the behavior
of the quantum vacuum is more certain. This provides a unique probe of the
physics of extra dimensions without the necessity of dealing with unknown
effects at energies larger than $M_*$.  This is particularly relevant in
brane-world scenarios where gravity propagates in a higher-dimensional space
while our visible Universe is a three-dimensional brane in the bulk of extra
dimensions \cite{Arkani-Hamed:1998rs}.

In this paper we initially assume a five-dimensional world where our visible
Universe is a three-dimensional brane located at a given point in the fifth
dimension. We consider the simplest possibility that inflation is a brane
effect, {\it i.e.,} it is driven by a scalar field living on our
three-dimensional brane, and study the effects of the transdimensional physics
on the spectrum of the primordial density perturbations produced during the
epoch of inflation.

Our findings indicate that despite the fact that the power spectrum of scalar
perturbations remains unchanged, the existence of the fifth dimension is
imprinted on the spectrum of gravitational waves generated during inflation.
The tensor spectrum receives a correction proportional to $(HR)^2$, where $H$
is the Hubble rate during inflation and $R$ is the size of the extra
dimension. Generalizing our results to the case of more than one extra
dimension and to warped geometries, we show that the numerical coefficient of
the correction term depends upon the details of the spacetime geometry of the
extra dimensions.  In four-dimensional single-field models of inflation there
exists a consistency relation relating the amplitude of the scalar
perturbations, the amplitude of the tensor perturbations, and the tensor
spectral index. We compute the correction to such a consistency relation from
transdimensional physics. Surprisingly enough, we find that at lowest order in
the slow roll expansion, the four-dimensional relation is quite robust and does
not suffer corrections from extra-dimensional physics, at least in not the
cases addressed in this paper.

Some similar conclusions have been reached in Refs.\ \cite{lmw,mwbh} for a
particular five-dimensional setup in which the expansion law on the brane has a
non-standard expression. Instead, we will focus on the case where the radius of
the extra dimension is stabilized, leading to an ordinary Friedmann law. So
any effect should be attributed to the non-trivial geometry along the extra
dimension, rather than any modified cosmology on the brane. However, the
formalism used in Ref.\ \cite{lmw} has many similarities with ours.

Our paper is organized as follows. In Section II we study the five-dimensional
background with a stabilized radius. In Section III we compute the power
spectrum of the tensor modes generated during inflation, while in Section IV we
calculate the power spectrum of scalar perturbations.  Section V is devoted to
the consistency relation and Section VI to a generalization of our findings to
more than one extra dimension and to warped geometries. Finally, in Section VII
we draw our conclusions.  The paper also contains an Appendix where we collect
the background and perturbed Einstein equations.

\section{A Five-dimensional background with a stabilized radion}

We consider a framework consisting of a $(3+1)$-dimensional brane embedded in a
five-dimensional bulk with a stabilized radius.  The coordinate along the extra
dimension is taken to be $0 \leq y < 2 \pi R$ (eventually, one may consider to
orbifold the circle by a $Z_2$-symmetry that identifies $y$ with $-y+2\pi R$
obtaining the segment $S^1/ Z_2$), and the brane is located at $y=0$ at zeroth
order in the perturbations.  Latin indices ($i=1,2,3$) label the ordinary three
space dimensions; Greek indices ($\mu=0,1,2,3,5$) run over time, the three
ordinary spatial dimensions, and the extra dimension $\mu=5$ (there is no
$\mu=4$).  The background metric may be taken to be of the form
\begin{equation}
ds^2 = n^2(t,y) \, dt^2 - a^2(t,y) \delta_{ij} d x^i \, d x^j - dy^2 .
\label{backgroundmetric}
\end{equation}
Through a redefinition of time we can always impose the condition $n(t,0)=1$ in
order to obtain the familiar equations on the brane, where the induced metric
is simply $ds^2 = dt^2 - a_0^2(t)\, \delta_{ij}\,dx^i\,dx^j$ \cite{footnotea0}.
In the background metric there is no time-dependent $b^2$ term multiplying
$dy^2$ because we assume that the radion is stabilized by some unknown
high-energy mechanism, and we are free to set $b^2=1$.  Then for consistency we
must assume that the bulk energy-momentum tensor has a non-vanishing $(55)$
component \cite{Kanti:2000rd} that accounts for the radion stabilizing
mechanism, while for simplicity we take the other components of the bulk
energy-momentum tensor to be zero. This situation can be achieved by
introducing a potential for the radion in the bulk that vanishes at the minimum
and whose mass parameter is much larger than the other relevant mass scales.
We will see in the following that our results can be generalized to cases with
a non-vanishing bulk cosmological constant and a brane tension (like in the
Randall--Sundrum framework \cite{rs}).

We suppose that the vacuum energy driving inflation is localized on our
three-brane at $y=0$ so that the brane energy-momentum tensor provided by the
inflaton brane-field $\varphi$ is of the form $T^\mu_{\ \ \nu}=\delta(y)\,{\rm
diag} (\rho,-p,-p,-p,0)$.  This might be considered the simplest
higher-dimensional scenario to investigate the effects of extra dimensions on
cosmological scales. Of course, one may envisage extensions of our set up, such
as assuming that the inflaton field $\varphi$ lives in the bulk made of one or
more than one extra dimension, or that the spacetime geometry is warped. We
will comment of these generalizations at the end of the paper.

We look for solutions of the background (unperturbed) Einstein equations
\cite{ex}:
\begin{eqnarray}
G^0_{\ 0} & = & 
\frac{3}{n^2} \left( \frac{\dot{a}}{a} \right)^2 - 3 \left[ \frac{a''}{a}
+ \left( \frac{a'}{a} \right)^2 \right] = M_*^{-3} \ \delta(y) \ \rho(t) ,
\label{g00} \\
G^i_{\ j} & = & \left\{
\frac{1}{n^2}\left[2 \frac{\ddot{a}}{a}+\frac{\dot{a}}{a}
\left(\frac{\dot{a}}{a}-
2\frac{\dot{n}}{n}\right)\right] -2\frac{a''}{a}
-\frac{a'}{a}\left(\frac{a'}{a}+2\frac{n'}{n}\right)-\frac{n''}{n}
\right\}\delta_{ij} \nonumber \\ 
& = & - M_*^{-3} \ \delta(y) \ p(t) \ \delta_{ij} ,
\label{gij} \\ 
G^0_{\ 5} & = & 
\frac{3}{n^2}  
\left( \frac{\dot{a}}{a} \frac{n'}{n} - \frac{\dot{a}'}{a} \right) = 0 ,
\label{g05}
\end{eqnarray}
where $M_*$ is the fundamental gravitational mass, $\rho$ is the energy density
on the brane, and $p$ is the pressure on the brane. An overdot denotes
derivation with respect to $t$, while a prime superscript denotes
differentiation with respect to $y$.  The $G^5_{\ 5}$ equation accounts for the
stabilization of the radion and provides a constraint on $T^5_{\ 5}$, not on
the metric.  Other components vanish at zero order in perturbations.  In
general, the only solution of Eqs.\ (\ref{g00})--(\ref{g05}) such that
$n(t,0)=1$ is easily found to be
\begin{eqnarray}
a(t,y) &=& \dot{a}(t,0) \left[ y^2 - 2 \pi R y
+ \frac{6 \pi R M_*^3}{\rho(t)} \right]^{1/2} , \nonumber \\
n(t,y) &=&  \frac{\dot{a}(t,y)}{\dot{a}(t,0)}.
\label{aandn}
\end{eqnarray}
Note that the solution for the background metric is automatically
$Z_2$-symmetric. Later, we will assume that this is also the case for metric
perturbations.

The expression for $a(t,y)$ leads to the standard Friedmann law on the brane
expected with a stabilized radion:
\begin{equation}
H^2 \equiv \left[\frac{\dot{a}(t,0)}{a(t,0)}\right]^2 = 
\frac{1}{2 \pi R M_*^3} \frac{\rho(t)}{3}.
\label{friedman}
\end{equation}

In general, the presence of matter on the brane will cause a small
readjustment of the radion with respect to its equilibrium value in vacuum.
This shift generates corrections to the Friedmann law in Eq.\ (\ref{friedman})
which are quadratic in $\rho$. Under our assumption that the radion is
stabilized by a bulk potential characterized by a mass much larger than
the other relevant energy scales, we can safely neglect these corrections.

From this expansion law we can define the four-dimensional gravitational
constant to be $M_{P}^2 \equiv (8 \pi G)^{-1} \equiv 2 \pi R M_*^3$.  Note that
when $\pi R H \geq 1$, the solution for the scale factor is singular: it is not
consistent to impose radion stability when the size of the extra dimension is
larger than the Hubble radius, $H^{-1}$. (If the Hubble radius is interpreted
as the causal horizon, this just means that the stabilization mechanism must
remain causal.)

The singular nature of the scale factor for $\pi R H \geq 1$ is not an issue
since we are only interested in the case in which the size of the extra
dimension is smaller than the Hubble radius, $\pi R H \leq 1$, and the
cosmological framework is expected to be almost described by four-dimensional
physics (up to the correction factors that we wish to calculate).

Therefore, we assume that the energy density during inflation is smaller than
$3 M_{P}^2 / (\pi R)^2$, and that deviations from the standard Friedmann law
are suppressed up to this scale.

Matching the discontinuity in the components of the Einstein equations
(\ref{g00}) and (\ref{gij}) gives the well known jump conditions for $a'$ and
$n'$:
\begin{equation}
\left[ \frac{a'}{a} \right]_{0}^{2 \pi R} = \frac{1}{3M_*^3}\ \rho , 
\qquad
\left[ \frac{n'}{n} \right]_{0}^{2 \pi R} = - \frac{1}{3M_*^3} \ 
(2 \rho + 3 p) ,
\end{equation}
where, for any function $f$, we define
\begin{equation}
[f]^\alpha_\beta \equiv
f(\alpha)-f(\beta)\, .
\end{equation}
The restriction of the equation for $G^0_{\ 5}$ on the brane yields the usual
energy conservation law for a perfect fluid: $\dot{\rho} + 3 H (\rho + p) = 0$.

When the brane only contains a homogeneous inflaton field $\bar{\varphi}(t)$
with potential $V$, the fluid energy conservation law gives the Klein--Gordon
equation: $\ddot{\bar{\varphi}} + 3 H \dot{\bar{\varphi}} + \partial V /
\partial \bar{\varphi} = 0$.

If we assume that the density $\rho$ is constant over time, the scale factor is
a separable function of time and $y$, and the brane undergoes de Sitter
expansion:
\begin{equation}
a(t,y) = a_0(t) n(y) , \qquad a_0(t) \propto \exp(H t) , \qquad
n(y) = \left[ H^2 (y^2 - 2 \pi R y) + 1 \right]^{1/2} . 
\label{desitter}
\end{equation}

\section{The Primordial spectrum of tensor perturbations \label{tensorsection}}

In this section we compute the present-day power spectrum of tensor modes
generated by a primordial period of inflation on our visible brane at $y=0$.

The tensor perturbation of the metric is defined, as usual, in terms of a
traceless transverse tensor $h_{ij}$ such that
\begin{equation}
ds^2 =  n^2(t,y) \, dt^2 - a^2(t,y) (\delta_{ij} + h_{ij}) 
d x^i \, d x^j - dy^2 .
\end{equation}
One learns from the perturbed Einstein equations that the two degrees of
polarization contained in $h_{ij}$ obey the wave equation
\begin{equation}
\ddot{h} + \left(3 \frac{\dot{a}}{a} - \frac{\dot{n}}{n} \right) \dot{h}
- \frac{n^2}{a^2} \Delta h 
- n^2 h'' - n^2 \left(3 \frac{a'}{a} + \frac{n'}{n} \right) h' = 0 ,
\label{eq-h}
\end{equation}
where $h$ is normalized in such a way that $h^{ij} h_{ij} = h^2/2$.
  
Note that a free scalar field propagating in the bulk would have the same
equation of motion as $h$.  In the de Sitter background defined in Eqs.\
(\ref{desitter}), and in a Fourier expansion with respect to the three spatial
coordinates $x_i$, the equation reads
\begin{equation}
\ddot{h}_k + 3 H \dot{h}_k + \frac{k^2}{a_0^2} h_k - n^2 h_k'' - 4 n'n \ h_k' 
= 0 .
\label{eq.h}
\end{equation}
We see from Eq.\ (\ref{desitter}) that $(n^2)'$ is not continuous on the brane,
and has a jump
\begin{equation}
\left[ 2nn' \right]_0^{2 \pi R} = 4 \pi R H^2 .
\end{equation}
This implies that $(n^2)''$ contains a $\delta$ function:
\begin{equation}
(n^2)'' = 2 H^2 \left[1 - 2 \pi R \ \delta(y) \right] .
\end{equation}
The solutions of the mode equation, Eq.\ (\ref{eq.h}), are separable in $t$
and $y$, so we can expand $h_k$ in a sum of Kaluza--Klein modes:
\begin{equation}
h_k= a_0^{-3/2} n^{-2}
\sum_p \chi_p(t) g_p(y) ,
\label{def.gn}
\end{equation}
where $\chi_p(t)$ and $g_p(y)$ satisfy the equations
\begin{eqnarray}
\ddot{\chi}_p + \left( \frac{k^2}{a_0^2} - \frac{9}{4} H^2 + \omega_p^2 \right)
\chi_p & = & 0 \label{eq-chin} , \\
n^2 g_p'' + \left[ - (n^2)'' + \omega^2_p \right] g_p & = & 0 . 
\label{eq-gn}
\end{eqnarray}
In the definition of Eq.\ (\ref{def.gn}), the factor $a_0^{-3/2} n^{-2}$ was
introduced just for simplicity so that Eqs.\ (\ref{eq-chin}) and (\ref{eq-gn})
contain no friction terms.  The equation for $g_p$ has to two independent
solutions, which are given in terms of the Gauss hypergeometric function
${_2}F_1$,
\begin{equation}
g_p = c_1\ {_2}F_1 \left( \frac{-1-b}{4},\frac{-1+b}{4},\frac{1}{2},-x^2
\right) +
c_2\ x\  {_2}F_1 \left( \frac{1-b}{4},\frac{1+b}{4},\frac{3}{2},-x^2
\right) ,
\end{equation}
where the parameters $b$ and $x$ are defined by
\begin{equation}
b \equiv \sqrt{9-4\frac{\omega_p^2}{H^2}}, \qquad
x\equiv \frac{(y-\pi R)H}{\sqrt{1-(\pi RH)^2}}.
\end{equation}
The metric continuity condition $g_p(0)=g_p(2 \pi R)$ eliminates the solution
odd with respect to $(y- \pi R)$, and fixes $c_2=0$.  The constant $c_1$ is
determined by the normalization condition of $g_p$.  We will set the
wave function normalization condition to be
\begin{equation}
\int_{0}^{2 \pi R} \! \! \! dy \ n^{-2} \left| g_p \right| ^2 = 1  .
\label{wfnorm}
\end{equation}
 Integrating Eq.\ (\ref{eq-gn}) in a neighborhood of the
brane leads to the jump condition for $g'$:
\begin{equation}
\left[ g_p' \right]_0^{2 \pi R} = 4 \pi R H^2 g_p(0) . 
\label{gpdisc}
\end{equation}
This condition is satisfied only for a discrete set of possible values of
$\omega_p$, determined by the equation \cite{eigenvalues}
\begin{equation}
{_2}F_1 \left( \frac{-1-b}{4},\frac{-1+b}{4},\frac{1}{2},-a
\right) = \frac{(a+1)(b^2-1)}{8}\  
{_2}F_1 \left( \frac{3-b}{4},\frac{3+b}{4},\frac{3}{2},-a
\right) ,
\label{eigenvalue}
\end{equation}
with $a$ given by
\begin{equation}
a \equiv \frac{(\pi R H)^2}{1-(\pi R H)^2}.
\end{equation}

As far as the time dependence is concerned, the solution of Eq.\
(\ref{eq-chin}) for $\chi_p$ is a Bessel function, and can be normalized to the
adiabatic vacuum inside the Hubble radius using as usual the positive frequency
condition and the canonical commutation relations.  To do so, we start from the
five-dimensional action
\begin{equation}
S =  \frac{1}{8} 
\int dt\ dy\ d^3\!x \ M^3_* n a^3 \left( n^{-2} \dot{h}^2 
- a^{-2} \delta^{ij} \partial_i h \partial_j h - {h'}^2 \right) .
\end{equation}
Note that the factor of $na^3$ simply comes from the term $\sqrt{-g}$.  In
Fourier space and with the de Sitter background the action is
\begin{equation}
S =  \frac{1}{8} \int dt\ dy\ d^3\!k \ M^3_* n^2 a_0^3 
\left[ \dot{h}_k \dot{h}^*_k 
+ \frac{k^2}{a_0^2} h_k h^*_k - n^2 h'_k {h^*_k}' \right] .
\end{equation}
Following Eq.\ (\ref{def.gn}), we can expand each mode along the basis formed
by the functions $g_p$.  We define
\begin{eqnarray}I_{mp} &=& \int_0^{2 \pi R} \! dy\ n^{-2} g_m \  g_p^*  ,
\nonumber \\
J_{mp} &=& \int_0^{2 \pi R} \! dy \left(g_m' - 2 \frac{n'}{n} g_m \right)
\left({g_p^*}' - 2 \frac{n'}{n} g_p^* \right) .
\end{eqnarray}
After integration over $y$, the effective action reads
\begin{eqnarray}
S & = &  \frac{1}{8} \int dt \ d^3\!k \  M_*^3 \sum_{m,p}
\left\{ \left[
\dot{\chi}_m \dot{\chi}^*_p  - \frac{3}{2} H 
(\chi_m \dot{\chi}^*_p  + \dot{\chi}_m \chi^*_p) 
+ \left(\frac{k^2}{a_0^2} + \frac{9}{4} H^2 \right) \chi_m \chi^*_p \right]
I_{mp}  \right. \nonumber \\
& & \left. \phantom{\left(\frac{k^2}{a_0^2}\right)}
- \chi_m \chi^*_p J_{mp} \right\} .
\end{eqnarray}
Even without knowing explicitly the expression for the $g_p$'s, we can find 
$I_{mp}$ and $J_{mp}$. The first intermediate step is to integrate by parts
\begin{equation}
\int_0^{2 \pi R} \! dy \ g_m \ {g_p^*}'' = - \int_0^{2 \pi R} \! dy \ 
g_m' \ {g_p^*}' + 4 \pi R H^2 g_m(0) \ g_p^*(0) .
\end{equation}
So, the integral on the left-hand side has the hermitian symmetry.
Then, we use Eq.\ (\ref{eq-gn}) and write
\begin{eqnarray}
g_m \ {g_p^*}'' + \left[- 2 H^2 \left[1 - 2  \pi R \ \delta(y) \right] 
+ \omega^2_p\right] n^{-2} g_m \ g_p^* 
&=& 0 , \nonumber \\
g_m'' \ g_p^* + \left[- 2 H^2 \left[1 - 2 \pi R \ \delta(y) \right]
+ \omega^2_m\right] n^{-2} g_m \ g_p^* &=& 0 .
\end{eqnarray}
We subtract these two equations and integrate over $y$, taking advantage of 
the previously found symmetry. We are left with
\begin{equation}
(\omega^2_p - \omega^2_m)  \int_0^{2 \pi R} \! dy \ n^{-2} g_m \ g_p^* = 0 .
\end{equation}
So, unless $\omega^2_p=\omega^2_m$, the above integral vanishes.  Given the
wave function normalization condition of Eq.\ (\ref{wfnorm}), we conclude that
$I_{mp} = \delta_{mp}$.  We can also integrate by parts
\begin{equation}
J_{mp} = \int_0^{2 \pi R} \!  dy \ \left[n^{-2} \left(g_m' - 
2 \frac{n'}{n} g_m \right) \right] \left( n^2 {g_p^*}' - 2 n' n g_p^* \right) .
\end{equation}
Using the equation of motion and the jump condition for $g_p$, we find $J_{mp}
= \omega_p^2 \delta_{mp}$.  So, the effective four-dimensional action is
diagonal:
\begin{equation}
S = \sum_{p}  \frac{1}{8} \int dt \ dk^3 \ M_*^3 
\left[ \dot{\chi}_p \dot{\chi}^*_p + \left( \frac{k^2}{a_0^2} 
+ \frac{9}{4} H^2 - \omega^2_p \right) \chi_p \chi^*_p - \frac{3}{2} H 
(\chi_p \dot{\chi}^*_p  + \dot{\chi}_p \chi^*_p) \right] .
\end{equation}
Each Kaluza--Klein mode $\chi_p$ has the same action as a free field in
four-dimensional de Sitter spacetime, and can be quantized following the
standard procedure. Namely, the adiabatic vacuum can be defined in the
sub-horizon limit $k/a_0 \gg H$ in which Minkowski spacetime is
asymptotically recovered.  Then the canonical commutation relation gives
\begin{equation}
\hat{p}_{\chi_p} =  \frac{1}{8} M_*^3 \dot{\chi}_p , \qquad
\left[ \hat{\chi}_p, \hat{p}_{\chi_p}^{\dagger} \right] = i,
\end{equation}
which leads to the Wronskian condition $\chi_p \dot{\chi}_p^* - \dot{\chi}_p
\chi_p^* = 8i/M_*^3$.

Let us focus on the zero mode. Equation (\ref{eq-gn}) with $\omega_p=0$ has the
obvious solution $c_0 n^2$, where $c_0$ is a constant of integration. This
solution is automatically continuous and satisfies the jump condition on the
brane. The constant of integration is obtained from the condition of Eq.\
(\ref{wfnorm}):
\begin{equation}
c_0 = \left( \int_0^{2 \pi R} \! dy \ n^2 \right)^{-1/2} 
= \left[ 2 \pi R \left(1 - \frac{2}{3} \pi^2 R^2 H^2\right) \right]^{-1/2} . 
\end{equation}
The function $\chi_0(t)$ is a Bessel function of index $3/2$ and has a simple
analytic expression (we retain only the positive frequency solution and we
normalize with the Wronskian condition above),
\begin{equation}
\chi_0 = \frac{2}{M_*^{3/2}} \sqrt{\frac{a_0}{k}} 
\left(i \frac{a_0H}{k} + 1 \right) \exp\left(i \frac{k}{a_0H}\right) .
\end{equation}

\begin{figure}[t]
\centerline{\epsfxsize=4.5in \epsfbox{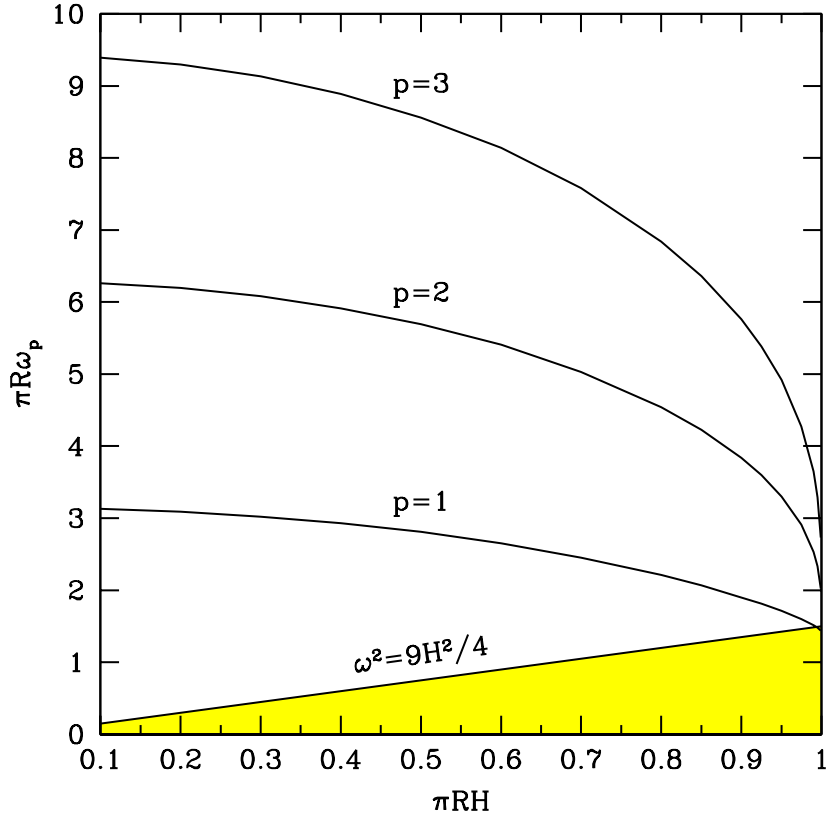}}
\caption{\label{fig} Effective mass $\omega_p$ of the first excited 
Kaluza--Klein modes for tensor perturbations.  For $p=1,2,3,$ we plot $\pi R
\omega_p$ as a function of $\pi R H$. Note that as expected, as $\pi R H$ 
approaches zero, $\pi R \omega_p$ approaches $p\pi$.}
\end{figure}

After horizon crossing, $|\chi_0|$ grows like $a_0^{3/2}$. The behavior of the
other Kaluza--Klein modes depends on the sign of $\omega_p^2 - 9H^2/4$.  As
shown in Fig.\ \ref{fig}, even $\omega_1^2$ is larger than $9H^2/4$
except for a marginal range when $\pi HR$ is very close to 1.  So, all massive
Kaluza--Klein modes oscillate at late time with a constant amplitude, and are
quickly suppressed with respect to the zero mode by the factor $a_0^{3/2}$.

In addition, the spectrum of the massive Kaluza--Klein modes is extremely blue,
as is usually the case for a scalar field with a mass larger than the Hubble
parameter during inflation. This means that for astronomical scales of
interest the contribution to the tensor spectrum from Kaluza--Klein modes is
practically zero.  So, we can focus on the asymptotic value of $h_k$ arising
from the zero-mode contribution
\begin{equation}
h_k(t,0) \rightarrow a_0^{- 3/2} \chi_0(t) g_0(0)  \rightarrow i 
\left[2 \pi R M_*^3 \left(1 - \frac{2}{3} \pi^2 R^2 H^2\right) \right]^{-1/2} 
\frac{2}{k^{3/2}} \ H . \label{ash}
\end{equation}
Even when the de Sitter stage ends, Eq.\ (\ref{eq-h}) shows that the
zero mode remains frozen on wavelengths larger than the horizon, as is the
case in four-dimensional physics. Therefore, the primordial spectrum of
gravitational waves at horizon re-entry is still given by Eq.\ (\ref{ash}),
where $H$ has to be evaluated at the time of the first horizon crossing during
inflation:
\begin{equation}
{\cal P}_T(k) \equiv \frac{k^3}{2\pi^2} \left|h_k(y=0)\right|^2 =
\frac{2}{\pi^2}\left(\frac{H(k)}{M_{P}}\right)^2 \frac{1}{1 - 2\pi^2 R^2 
H^2(k)/3}=\frac{\left. {\cal P}_T(k)\right|_{4D}}{1 - 2\pi^2 R^2 H^2(k)/3} ,
\label{psh}
\end{equation}
where $H(k)$ indicates the value of the Hubble parameter when a given
wavelength $\lambda=2\pi/k$ crosses the horizon, {\it i.e.}, when $k=a_0H$.
The power spectrum of tensor perturbations is normalized such that in a
critical density universe the energy density (per octave) in gravitational
waves, $\Omega_g(k)$, is related to ${\cal P}_T(k)$ in terms of the transfer
function, $T_g^2(k)$, by (the transfer function is discussed in Ref.\ \cite{G})
\begin{equation}
\Omega_g(k) = \frac{1}{24} T_g^2(k) {\cal P}_T(k) . 
\end{equation}  
The power spectrum of tensor modes is therefore enhanced compared to the
four-dimensional result by a factor $(1 - 2\pi^2 R^2 H^2/3)^{-1}$.

This correction factor has a simple explanation: it originates from the
zero-mode wave function normalization. However, it can be understood also in
terms of the effective gravitational Planck mass $\left.M_{P}^2\right|_{I}$
{\it during inflation}, defined by integrating the zero-mode action over
$y$. Indeed, the zero mode, $h_0 \propto n^{-2} g_0$, is constant along $y$ (as
expected for a free field with no source localized on the brane). So,
integrating the zero-mode action gives a factor (setting $a_0=1$ to isolate the
gravitational coupling) \cite{effective}
\begin{equation}
\left.M_{P}^2\right|_{I}= M_*^3
\int_0^{2 \pi R} \! \! dy \ \sqrt{- g} \ g^{00} =  M_*^3 \int dy \ n^2
=  M_{P}^2\left(1 - \frac{2}{3} \pi^2 R^2 H^2\right) .
\label{efpl}
\end{equation}
This shows that the enhancement of the tensor power spectrum can be rephrased
as a shift in the effective gravitational constant during inflation, when it is
defined from the effective gravitational action rather than from the expansion
law. The tensor perturbation is a purely five-dimensional field, while the
expansion law obtains from a density localized on the brane.  This implies that
the latter depends on the value of $n^2$ on the brane, not on its average.
Physically what happens is that the vacuum energy density present on our
visible brane during inflation warps the spacetime geometry in the bulk. This
effect is manifest in the nontrivial shape of the functions $a^2(y)$ and
$n^2(y)$ during the inflationary epoch. As a result, the graviton zero mode,
which is free to spread out in the bulk, feels a smaller Planck mass during
inflation. Therefore, today we receive a flux of gravitational waves
primordially generated during inflation which is larger than its
four-dimensional counterpart because during inflation gravity was stronger.

\section{The Primordial spectrum of scalar perturbations}

In this section we compute the present-day power spectrum of scalar modes
generated by a primordial period of inflation on our visible brane at $y=0$.

The first-order scalar perturbations of the metric can be expressed 
as \cite{perturb}
\begin{eqnarray}
ds^2 & = & n^2 (1 + 2 \phi) \, dt^2 
- a^2 [(1 - 2 \psi)\delta_{ij} + 2 \partial_i \partial_j E ] \, d x^i \, d x^j
+ 2 \partial_i B \, d x^i \, d t \nonumber \\
& &
+ 2 \partial_i w \, dx^i \, dy + 2 \delta g_{05} \,dt \, dy
- (1 - \delta g_{55}) \, dy^2  .
\end{eqnarray}
The perturbed brane position is specified by another function, $\delta y (t,
x^i)$.  

Five-dimensional gauge transformations of the form $x^{\mu} \rightarrow x^{\mu}
+ \xi^{\mu}$, where $\xi^{\mu} = (\xi^0, \partial^i \xi, \xi^5)$, induces the
transformations
\begin{eqnarray}
\phi &\rightarrow& \phi + \dot{\xi}^0 + \frac{\dot{n}}{n} \xi^0 
+ \frac{n'}{n} \xi^5  , \nonumber \\
\psi &\rightarrow& \psi - \frac{\dot{a}}{a} \xi^0 - \frac{a'}{a} \xi^5 ,
\nonumber \\
E &\rightarrow& E + \xi ,\nonumber \\
B &\rightarrow& B + n^2 \xi^0 - a^2 \dot{\xi}  , \nonumber \\
w &\rightarrow& w + \xi^5 + a^2 \xi' , \nonumber \\
\delta g_{05} &\rightarrow&  \delta g_{05} + \dot{\xi}^5 - n^2 {\xi^0}' ,
\nonumber \\
\delta g_{55} &\rightarrow&  \delta g_{55} + 2 {\xi^5}' ,
\end{eqnarray}
while the new brane position is $\delta y (t, x^i) 
+ \xi^5(t,x^i,\delta y (t, x^i))$.

We will work in a particular gauge, the Gaussian normal gauge.  The same gauge
choice was made in {\it e.g.,} Ref.\ \cite{gauge}.  In order to eliminate
$\delta g_{55}$, we choose $2 {\xi^5}' = - \delta g_{55}$. This fixes the
function $\xi^5(t, x^i, y)$ up to a boundary condition, {\it i.e.,} up to an
arbitrary function of $(t, x^i)$ on one hypersurface (for instance, on the
brane).  The most convenient boundary condition is $\xi^5(t,x^i,\delta y (t,
x^i)) = - \delta y (t, x^i)$, in order to shift the brane position to $y=0$,
even at first order in perturbations.  Similarly, in order to eliminate $w$, we
may choose $a^2 \xi' = - w - \xi^5$, with the boundary condition $\xi = - E$ on
the brane in order to have $E=0$ on the brane.  Finally, in order to eliminate
$\delta g_{05}$, we choose $n^2 {\xi^0}' = \delta g_{05} + \dot{\xi}^5$, with
the boundary condition $n^2 \xi^0 = - B + a^2 \dot{\xi}$, so that $B$ also
vanishes on the brane.  Of course $E$ and $B$ are still non-zero in the bulk.
The perturbed metric reduces to
\begin{equation}
ds^2 = n^2 (1 + 2 \phi) \, dt^2 
- a^2 [(1 - 2 \psi) \delta_{ij} + 2 \partial_i \partial_j E ] \, 
d x^i \, d x^j
+ 2 \partial_i B \, d x^i \, d t - dy^2 .
\end{equation}
The induced metric on the brane is diagonal, and involves only the
perturbations $\phi_0$ and $\psi_0$.  It is identical to the four-dimensional
perturbed metric in the so-called longitudinal gauge.  Since the system is
symmetric in $y \longleftrightarrow 2 \pi R - y$, the perturbations are
expected to be even functions with respect to $(y - \pi R)$.  In the following,
the terms ``even'' and ``odd'' will be meant always with respect to $(y - \pi
R)$.

We give in Eqs.\ (\ref{Einstein.00})--(\ref{Einstein.i5}) the expression of the
perturbed Einstein equations in the Gaussian normal gauge.  In general, the
restrictions of the $G^0_{\ 5}$ and $G^i_{\ 5}$ equations on the brane provide
the continuity and Euler equations.  When the brane contains only a perturbed
scalar field $\varphi (t,x^i) = \bar{\varphi}(t) + \delta \varphi (t,x^i)$, the
$\delta G^i_{\ 5}$ equation is trivially satisfied, while the $\delta G^0_{\
5}$ equation gives the standard perturbed Klein--Gordon equation:
\begin{equation}
\delta \ddot{\varphi} + 3 \frac{\dot{a}_0}{a_0} \delta \dot{\varphi}
+ \left( \frac{\partial^2 V}{\partial \varphi^2} 
- \frac{\Delta}{a_0^2} \right) \delta \varphi
= \dot{\bar{\varphi}} (\dot{\phi}_0 + 3 \dot{\psi}_0) - 2 
\frac{\partial V}{\partial \varphi} \phi_0 .
\label{pert-KG}
\end{equation}
The other components of the perturbed Einstein equations contain some second
derivatives with respect to $y$ that have to be matched with source terms on
the brane. However, to first order in the perturbations, the scalar field
cannot generate anisotropic stress on the brane: $\delta T_{ij}$ is
proportional to $\delta_{ij}$.  This imposes the continuity of $E''$ across the
brane, and therefore, since $E'$ is odd, $E'_0=0$. The other perturbations are
sourced on the brane and have to satisfy the jump conditions
\begin{eqnarray}
-3 \left[\psi'\right]_0^{2\pi R} & = & M_*^{-3} \delta T^0_{\ 0}
= \dot{\bar{\varphi}} \ \delta \dot{\varphi} - \dot{\bar{\varphi}}^2 \phi
+ \frac{\partial V}{\partial \varphi} \delta \varphi , \nonumber \\
- \left[\phi'\right]_0^{2\pi R} + 2 \left[\psi'\right]_0^{2\pi R} & = & 
     - M_*^{-3} \delta T^i_i
= \dot{\bar{\varphi}} \ \delta \dot{\varphi} - \dot{\bar{\varphi}}^2 \phi
- \frac{\partial V}{\partial \varphi} \delta \varphi , \nonumber \\
- \frac{1}{2} \left[B'\right]_0^{2\pi R} 
       & = & M_*^{-3} \dot{\bar{\varphi}} \ \delta \varphi . 
\end{eqnarray}

\subsection{A master equation for the scalar perturbations}

We would like to find an equation of motion for a single variable that would
account for the full scalar perturbation dynamics, as Eq.\ (\ref{eq-h}) did
for tensor perturbations.  Such a master equation has already been found in the
case of a maximally-symmetric background spacetime \cite{muk}, but not in
cases where $T_{55}$ accounts for the radion stabilization.  The best approach
is to work with a set of variables reflecting some gauge-invariant quantities.
By studying 5-dimensional gauge transformations, it is straightforward to show
that one can build four independent gauge-invariant quantities out of the full
set of scalar perturbations of the metric. In our gauge, these quantities
reduce to
\begin{eqnarray}
\Psi_{\phi} & = & \phi - \frac{\dot{B}}{n^2} 
+ a_0^2 [ - \ddot{E} - 2 H \dot{E} + n'n E'] , \nonumber \\
\Psi_{\psi} & = & \psi + H \frac{B}{n^2} 
+ a_0^2 [ H \dot{E} - n'n E'] , \nonumber \\
\Psi_{05} & = & B' - 2 \frac{n'}{n} B 
+ 2 a_0^2 n^2 [ \dot{E}' + H E' ] , \nonumber \\
\Psi_{55} & = & 2 a_0^2 [ n^2 E'' + 2 n'n E' ] .
\end{eqnarray}
Similarly, it is possible to build a gauge-invariant quantity out of the scalar
field perturbation $\delta \varphi$ and the metric perturbations $E$ and $B$. 
However, in our gauge $E$ and $B$ vanish on the brane, so $\delta
\varphi$ directly reflects the gauge-invariant field perturbation.
Although the Einstein tensor is not gauge invariant, some of its components can
be expressed in terms of  $(\Psi_{\phi}, \Psi_{\psi}, \Psi_{05}, \Psi_{55})$.
We write the Einstein
equations in the de Sitter background, first in terms of $(\phi, \psi, E, B)$
(see Eqs.\ (\ref{Einstein.DeSitter.00})--(\ref{Einstein.DeSitter.i5}) of the
Appendix), and then in terms of the above variables (some terms in $E$ and
$E'$ still remain).  The traceless part of $\delta G^i_{\ j}$ just gives
\begin{equation}
\Psi_{\psi} - \Psi_{\phi} - \frac{1}{2} \Psi_{55} = 0 ,
\label{traceless}
\end{equation}
and allows us to eliminate easily one of the four variables: instead of
($\Psi_{\psi}$, $\Psi_{\phi}$, $\Psi_{05}$, $\Psi_{55}$), we can work with
($\Sigma=\Psi_{\psi} + \Psi_{\phi}, \Psi_{05}, \Psi_{55}$). Then, the equation
for $\delta G^i_{\ 0}$ provides a simple relation between $\Sigma$ and
$\Psi_{05}$:
\begin{equation}
2n^2 \left(\dot{\Sigma} + H \Sigma\right) + \left(n^2 \Psi_{05}\right)' 
= 2 M_*^{-3} \dot{\bar{\varphi}} \delta \varphi \ \delta(y) .
\label{rel.Sigma.05}
\end{equation}
The equation for $\delta G^0_{\ 0}$, combined with the previous one,
gives a relation between $\Sigma$ and $\Psi_{55}$:
\begin{eqnarray}
& & \frac{3}{2} \left( n^2 \Sigma'' + 4 n' n \Sigma' \right) 
+ 3 H (H \Sigma + \dot{\Sigma} ) + \frac{\Delta}{a_0^2} \Sigma
+ \frac{3}{2} \left( \frac{n^2}{2} \Psi_{55}'' + 3 n'n \Psi_{55}'
+ 3 H^2 \Psi_{55} \right) \nonumber \\
& & \ \ \ \ \ \ = M_*^{-3} \left[ \delta T^0_{\ 0} + 6 H \dot{\bar{\varphi}} 
\delta \varphi + 3 H^2 \ \Psi_{55}(y=0) \right] \delta(y) .
\label{rel.Sigma.55}
\end{eqnarray}
Finally, the system is closed, for instance, by the equation for $\delta G^i_{\
5}$. Indeed, the quantity $a_0^2 n^{-2} (n^4 \delta G^i_{\ 5})'=0$ can be
combined with the previous constraints in Eqs.\ (\ref{rel.Sigma.05}) and
(\ref{rel.Sigma.55}), to lead to a master equation for $\Sigma$:
\begin{eqnarray}
\ddot{\Sigma} - H \dot{\Sigma} - \frac{\Delta}{a_0^2} \Sigma
- n^2 \Sigma'' - 4 n'n \Sigma' - 2 H^2 \Sigma
& = & M_*^{-3} \left[ \delta T^i_{\ i} - \delta T^0_{\ 0} + 2 a_0
\left( \frac{\dot{\bar{\varphi}} \delta \varphi}{a_0} \right)^. \right]
\delta(y) \nonumber \\
&=& 2 M_*^{-3} \left( - H \dot{\bar{\varphi}} \delta \varphi 
+ \ddot{\bar{\varphi}} \delta \varphi + \dot{\bar{\varphi}}^2 \phi_0 
\right) \delta(y) .
\label{master}
\end{eqnarray}
In the bulk, this equation looks like a five-dimensional wave equation, and in
terms of the rescaled variable $a_0^{-2} \Sigma$, it would be identical to that
of the tensor perturbations or to that of a canonically normalized bulk scalar
field.  The first difference with the tensor case is the presence of a source
localized on the brane, which imposes a jump condition on the derivative
$\Sigma'$:
\begin{equation}
 - \frac{1}{2} \left[\Sigma'\right]_0^{2\pi R} 
= M_*^{-3} \left( H \dot{\bar{\varphi}} \delta \varphi 
- \ddot{\bar{\varphi}} \delta \varphi - \dot{\bar{\varphi}}^2 \phi_0 \right) .
\label{Sigma.jump}
\end{equation}
A similar condition would be found in the case of a bulk scalar field sourced
on the brane.  However, there is a second difference, reflecting the
complicated structure of the Einstein equations, and the integro-differential
relations between the various perturbations and boundary conditions.  By
integrating Eq.\ (\ref{rel.Sigma.05}) over the circle $0 \leq y \leq 2 \pi R$,
we get an integrability condition for $\Sigma$:
\begin{equation}
\{ H + \partial_t \} \int_0^{2 \pi R} \! \! \! dy \ n^2 \Sigma =
M_*^{-3} \dot{\bar{\varphi}} \delta \varphi  .
\label{Sigma.integ}
\end{equation}

Any even solution of the master equation, Eq.\ (\ref{master}), matching the
jump and integrability conditions, Eqs.\ (\ref{Sigma.jump}) and
(\ref{Sigma.integ}), provides a solution of the full Einstein equations. At any
time one can compute the perturbations $(\phi_0, \psi_0)$ on the brane.
Indeed, the integral of Eq.\ (\ref{rel.Sigma.55}) over $y$ gives
$\Psi_{55}(y=0)$ as a function of $\Sigma$ and of the scalar field (remembering
that $\left[\Psi_{55}'\right]_0^{2\pi R}$ is given by the jump conditions,
while $\int_0^{2 \pi R} dy \Psi_{55} = 2 a_0^2 \left[n^2 E'\right]_0^{2\pi
R}=0$):
\begin{equation}
2 \pi R H^2 \ \Psi_{55} (y=0) =
- \left\{ \partial_t^2 + H \partial_t - \frac{\Delta}{3 a_0^2} \right\}
\int_0^{2 \pi R} \! \! \! \! \! dy \ \Sigma \
+ \frac{2}{3} M_*^{-3} \left( \dot{\bar{\varphi}} \delta \dot{\varphi}
+ 2 \ddot{\bar{\varphi}} \delta \varphi + 2 \dot{\bar{\varphi}}^2
\phi_0 \right) .
\label{Psi.55.0}
\end{equation}
Finally, the perturbations on the brane may be found from 
\begin{equation}
\psi_0 = \frac{1}{2} \left[ \Sigma (y=0)
+ \frac{1}{2} \Psi_{55} (y=0) \right] , \qquad
\phi_0 = \frac{1}{2} \left[ \Sigma (y=0) 
   - \frac{1}{2} \Psi_{55} (y=0) \right] .
\end{equation}

\subsection{Comparison to the four-dimensional solution}

In four dimensions, the absence of anisotropic stress implies $\phi_0=\psi_0$
and scalar metric perturbations are described by a single variable (matching
the Newtonian gravitational potential inside the Hubble radius). The exact
computation of the scalar power spectrum for scales leaving the Hubble radius
during inflation can be performed by various methods that take into account the
coupling between the metric and the scalar field perturbations
\cite{reviewrocky}.  For instance, one can integrate the equation of
propagation of the Mukhanov variable which is a combination of $\delta
\varphi$ and $\phi_0$ \cite{brand}. A second possibility is to solve a pair of
coupled equations: the perturbed Klein--Gordon equation (\ref{pert-KG}), and
one of the Einstein equations, for instance that for $\delta G^i_{\ 0}$:
\begin{equation}
\dot{\phi}_0 + H \phi_0 = \frac{\dot{\bar{\varphi}} \ \delta \varphi}{2 M_P^2}.
\label{4D.metric.constraint}
\end{equation}
In the present work, we are interested in the generalization of the scalar
power spectrum calculation only at the leading order in slow-roll parameters.
In four dimensions, such a calculation is done consistently by writing all the
equations of propagation in the exact de Sitter background; simultaneously, in
the expression for the perturbations one keeps the leading-order term in the
expansion parameter $\dot{H}/H^2$ [or equivalently $\dot{\bar{\varphi}}^2/(H^2
M_{P}^2)$], treated as a constant.  In that case, the sourcing of the field
perturbation by the metric perturbation [described by the right-hand side in
the Klein--Gordon equation Eq.\ (\ref{pert-KG})] can be safely neglected, and
one can solve the homogeneous equation
\begin{equation}
\delta \ddot{\varphi} + 3 H \delta \dot{\varphi}
+ \left( \frac{\partial^2 V}{\partial \varphi^2} 
+ \frac{k^2}{a_0^2} \right) \delta \varphi = 0 .
\label{free.pert}
\end{equation}
At leading order in slow-roll parameters we can also neglect $|V''| \ll H^2$
and write the solution, normalized to the adiabatic vacuum inside the Hubble
radius, as
\begin{equation}
\delta \varphi = \frac{1}{a_0 \sqrt{2 k}} \left( 1 + i \frac{H a_0}{k} \right)
\exp\left(i \frac{k}{a_0 H}\right) .
\end{equation}
The metric perturbation is just following the field evolution, according to
Eq.\ (\ref{4D.metric.constraint}). This gives in the two limits
(sub-Hubble-radius and super-Hubble-radius scales):
\begin{eqnarray}
\delta \varphi & \simeq & \frac{1}{a_0 \sqrt{2k}} \
\exp\left(i \frac{k}{a_0 H}\right) , \quad
\phi_0 =  \frac{i \dot{\bar{\varphi}}}{2 M_P^2 \sqrt{2 k^3}} \
\exp\left(i \frac{k}{a_0 H}\right) \quad
\left( \textrm{for}\,\,  \frac{k^2}{a_0^2} \gg H^2 \right) , \label{4d.sw} \\
\delta \varphi & \simeq & \frac{H}{\sqrt{2k^3}} ,
\hspace*{90pt} \phi_0 = \frac{\dot{\bar{\varphi}}}{2 M_P^2 \sqrt{2k^3}} 
\hspace*{81pt} \left( \textrm{for}\,\, \frac{k^2}{a_0^2} \ll H^2 \right) .
\end{eqnarray}
In both limits, we see that the right-hand side in the Klein--Gordon equation
is always very small, and acts as a source for the field perturbations only at
a sub-leading order in $\dot{H}/H^2$:
\begin{equation}
\left| \dot{\bar{\varphi}} (\dot{\phi}_0 + 3 \dot{\psi}_0) - 2 
\frac{\partial V}{\partial \varphi} \phi_0 \right| \sim 
\left| \frac{\dot{\bar{\varphi}}^2}{M_P^2} \delta \varphi \right|
\sim | \dot{H} \delta \varphi | \ll  H^2 |\delta \varphi| .
\end{equation}
Another way to reach the same conclusion is to look at the Mukhanov variable
$\xi= a_0 ( \delta \varphi + \dot{\bar{\varphi}}\phi_0/H)$.  It is
straightforward to show that at leading order in slow-roll parameters, $\delta
\varphi$ and $\xi/a_0$ are equal and share the same equations of propagation.
The distinction becomes only relevant when slow roll is marginally or
temporarily violated, for instance, in inflationary models with phase
transitions.

We reviewed this point in the four-dimensional case because it is crucial for
the five-dimensional calculation where we will also assume that at leading
order $\delta \varphi$ can be treated as a free field, obeying Eq.\
(\ref{free.pert}).  We will check {\em a posteriori} that the metric
perturbations just follow the field and do not alter $\delta \varphi$ in a
significant way.

\subsection{Long-wavelength solution in the five-dimensional model}

If the assumption that the scalar metric perturbations just follow the scalar
field perturbations is correct, then it is sufficient to study the coupling
between these two degrees of freedom in the long-wavelength regime in order to
know the power spectrum $k^3 |\phi_0|^2$ on super-Hubble-radius scales during
inflation. This is our purpose in this section.  In the next section we will
study the short-wavelength solution for consistency.

On super-Hubble-radius scales the scalar field perturbations are approximately
constant in time: $\delta \varphi = H / \sqrt{2 k^3}$ (we dropped the arbitrary
phase). We first look for a particular solution of the inhomogeneous master
equation, Eq.\ (\ref{master}), neglecting the Laplacian term. In slow roll only
the first of the three contributions to the source term on the right-hand side
is relevant because $|\ddot{\bar{\varphi}} \ \delta \varphi| \ll | H
\dot{\bar{\varphi}} \ \delta \varphi|$ and $ |\dot{\bar{\varphi}}^2
\phi_0 | \sim | \dot{H} \ \phi_0 | \ll H^2 |\phi_0|$.
A particular solution is found to be
\begin{equation}
\Sigma_i = \frac{1}{2 \pi R M_*^3} 
\frac{\dot{\bar{\varphi}} \delta \varphi}{H n^2} .
\label{part}
\end{equation}
This solution is constant in time, but not in $y$.  It matches both the jump
and integrability conditions Eqns.\ (\ref{Sigma.jump}) and (\ref{Sigma.integ})
(still at leading order in $\dot{H}/{H^2}$).  We are free to add to Eq.\
(\ref{part}) a solution of the homogeneous equation, {\it i.e.}, any even
solution $\Sigma_h$ of the sourceless master equation
\begin{equation}
\ddot{\Sigma}_h - H \dot{\Sigma}_h
- n^2 \Sigma_h'' - 4 n'n \Sigma_h' - 2 H^2 \Sigma_h = 0
\end{equation}
such that $\left[\Sigma_h'\right]_0^{2\pi R}=0$ and
\begin{equation}
\{ H + \partial_t\} \int_0^{2 \pi R} \! \! \! \! \! dy \ n^2 \ \Sigma_h = 0  .
\label{Sigma.hom.integ}
\end{equation}

We almost already have the solutions because the homogeneous master equation
written in terms of $(a_0^{-2} \Sigma_h)$ is the same as the equation for the
tensor modes. So $\Sigma_h$ is a sum of separable solutions (the Kaluza--Klein
modes) with the same values of $\omega_p$ as for the tensors, but with a
different normalization condition, Eq.\ (\ref{Sigma.hom.integ}).  For the zero
mode with $\omega_p=0$, the solution reads $(\Sigma_{h})_{p=0} = C_1 a_0^2 +
C_2 a_0^{-1}$, but the normalization condition imposes $C_1 = 0$. For heavy
modes with $\omega_p \geq 3H/2$, one gets on super-Hubble-radius scales
\begin{eqnarray}
(\Sigma_{h})_p =C_p a_0^{1/2} \exp\left(\pm i t \ \sqrt{\omega_p^2 - 
\frac{9}{4} H^2}
\right) \ n^{-2}(y) g_p(y) , 
\end{eqnarray}
where $g_p$ is an even solution of Eq.\ (\ref{eq-gn}). But, since 
the integral of
$g_p$ over $y$ does not vanish, the integrability condition imposes
$C_p=0$: in the limit under consideration, the Einstein equations are not
compatible with any significant contribution of heavy Kaluza--Klein modes.
The final solution reads
\begin{equation}
\Sigma = \frac{1}{2 \pi R M_*^3} 
\frac{\dot{\bar{\varphi}} \delta \varphi}{H n^2} + \frac{C_2}{a_0} ,
\label{sixtyfour}
\end{equation}
and corresponds to the usual combination of a growing adiabatic mode driven by
the scalar field, and a decaying mode that could be normalized only if the full
solution was known (from inside the horizon). Since the decaying mode becomes
rapidly negligible it will not concern us. Then, we can compute
$\Psi_{55}(y=0)$
 using Eq.\ (\ref{Psi.55.0}). The
leading-order terms are
\begin{equation}
2 \pi R H^2 \ \Psi_{55}(y=0)
\simeq - H \partial_t \int_0^{2 \pi R} \! \!  dy \ \Sigma +
\frac{2}{3} M_*^{-3} \dot{\bar{\varphi}} \delta \dot{\varphi} \simeq
-\frac{2}{3} \pi R \dot{H} \ \Sigma (y=0) \left[ 1 + {\cal O}(H^2 R^2) \right].
\label{eq:leading}
\end{equation}
So, to first order in $\epsilon = -\dot{H}/{H^2} = M_{P}^2 (V'/V)^2/2$, we
recover $\psi_0=\phi_0 = \Sigma_0/2$, and the metric perturbations on the brane
match exactly the four-dimensional result:
\begin{equation}
\phi_0 = \psi_0 = \frac{1}{4 \pi R M_*^3} 
\frac{\dot{\bar{\varphi}}}{H} \delta \varphi =
\frac{\dot{\bar{\varphi}}}{2 M_P^2 \sqrt{2k^3}}.
\end{equation}

Since our results indicate that on long wavelengths the perturbations of the
gravitational potential coincides with the four-dimensional one, the power
spectrum of scalar curvature perturbations will be given by the usual result
(we recall that during 
the de Sitter stage, the large wavelength
metric perturbations are related to the curvature
perturbations by a factor $\epsilon$) 
\begin{equation}
{\cal P}_S (k) = \frac{k^3}{2 \pi^2} \frac{\phi_0^2}{\epsilon^2} =
\frac{1}{8\pi^2}\frac{1}{\epsilon}\left(\frac{H(k)}{M_{P}}\right)^2 .
\label{cur}
\end{equation}
Here ${\cal P}_S(k)$ is defined in
terms of the observable power spectrum $P(k)$ and the scalar transfer function
$T^2(k)$ by
\begin{equation}
\frac{k^3}{2\pi^2}P(k) = \left(\frac{k}{aH}\right)^4 T^2(k) {\cal P}_S(k) .
\end{equation}  

In Eq.\ (\ref{cur}), the limit $\epsilon=0$ is singular as in four dimensions.
This corresponds to the exact de Sitter limiting case, for which
$\dot{\bar{\varphi}}=\ddot{\bar{\varphi}}=0$. Then, the master equation Eq.\
(\ref{master}) and the constraint equations Eqs.\
(\ref{Sigma.jump},\ref{Sigma.integ}) have vanishing right-hand sides.  The
single solution for $\Sigma$ at large wavelength is the decaying mode,
$C_2/a_0$ found in Eq.\ (\ref{sixtyfour}). Plugging this mode into Eq.\
(\ref{Psi.55.0}) shows that $\Psi_{55}(y=0)=0$ and $\phi_0 = \psi_0 = C_2 / 2
a_0 $. We reach the same conclusion as in four dimensions: for exact de Sitter
expansion, the scalar metric perturbations do not have a non-decaying solution
at large wavelength.

It is appropriate to re-emphasize our result that the scalar spectrum is
unaltered is only true to lowest order in the slow-roll parameters.  For
instance, one sees that Eq.\ (\ref{eq:leading}) has corrections of order
$(H^2R^2)$, but they are multiplied by $\epsilon$, so to lowest order in the
slow-roll parameters we can ignore them.

The absence of any correction factor in $(RH)^2$ can be interpreted in the
following way: Unlike the tensor degrees of freedom, which are five-dimensional
free fields quantized {\em in the bulk,} the scalar metric perturbations only
follow the scalar field. The later is quantized {\em on the brane} and has the
same behavior as in the four-dimensional case in all regimes.  So we only need
to study the coupling between the field and the metric in the long-wavelength
regime when the metric evolves as in the four-dimensional theory.  Moreover,
the coupling is localized on the brane, so that no signature remains from the
non-trivial geometry in the bulk.

\subsection{Short-wavelength solution in the five-dimensional model}

In the short-wavelength limit, the scalar field perturbations can be
approximated by $\delta \varphi = (a_0 \sqrt{2k})^{-1} \exp[-i(k/a_0)t]$, and
$k/a_0$ is a slowly varying parameter ($\delta \dot{\varphi} = - i (k/a_0)
\delta \varphi$).  We write the master equation with $H=0$ (and accordingly,
with $n=1$):
\begin{equation}
\ddot{\Sigma} - \frac{\Delta}{a_0^2} \Sigma - \Sigma'' = 0  .
\label{master.short}
\end{equation}
The brane source term that was proportional to $H$ has disappeared since the
jump in $\Sigma'$ across the brane is found to be negligible is the limit in
which the matter perturbations on the brane behave like a fluid with sound
speed $c_s^2 = -1$ ({\it i.e.}, $\delta T^0_{\ 0} = - \delta T^i_{\ i}$). Since
$\Sigma'$ is odd, this implies $[\Sigma']_0^{2\pi R} =0$.  The most general
even solution of Eq.\ (\ref{master.short}) can be written as a
Kaluza--Klein expansion (with $p=0,1,...,\infty$):
\begin{equation}
\Sigma =  \sum_{p = - \infty}^{+ \infty}  \Sigma_p 
=  \sum_{p = - \infty}^{+ \infty} 
 \left[ c_p \exp(i \nu_p t) + 
d_p \exp(- i \nu_p t) \right] \cos \left[\omega_p (y- \pi R) \right]  ,
\end{equation}
\begin{equation}
 \nu_p \equiv \left( \frac{k^2}{a_0^2} + \omega_p^2 \right)^{1/2} ,
\end{equation}
where $(c_p, d_p)$ are constants of integration and the $\omega_p$'s are
imposed by the continuity of $\Sigma'$:
\begin{equation}
\omega_p = \frac{p}{R} , \qquad p=0,1,...,\infty  .
\end{equation}
Inserting the general solution into the integrability condition Eq.\
(\ref{Sigma.integ}), which now reduces to
\begin{equation}
\int_0^{2 \pi R} \!  dy  \dot{\Sigma}
= M_*^{-3} \dot{\bar{\varphi}} \delta \varphi ,
\end{equation}
gives the two constraints
\begin{equation}
c_0=0, \quad d_0 = \frac{1}{2 \pi R M_*^3} 
\frac{i \dot{\bar{\varphi}}}{\sqrt{2 k^3}}.
\end{equation}
This implies
\begin{equation}
\Sigma_{p=0} = \frac{1}{2 \pi R M_*^3} \
i \frac{a_0}{k} \dot{\bar{\varphi}} \delta \varphi ,
\label{sigma.0}
\end{equation}
while the constants of integration for $p \geq 1 $ remain arbitrary.  In other
words, the zero mode is driven by the scalar field, as would be the case for
$\phi_0$ in four dimensions, while the heavy Kaluza--Klein modes are
independent of the matter on the brane.  So, the quantization should be done
first for the zero mode and the scalar field together since they only represent
one independent degree of freedom, then for each heavy Kaluza--Klein mode
separately following the same procedure as for the tensor modes (normalization
to the adiabatic vacuum). 

Let us focus on the quantization of the zero mode plus the field, since we know
that the heavy Kaluza--Klein modes will decouple when $kR/a_0 \ll 1$.  We first
must determine whether it is consistent to assume that the metric zero-mode
contribution to the perturbed Klein--Gordon equation is very small, and that
the field perturbation can be quantized as a free field, while the metric
zero-mode just follows.  In order to discover the answer, we need to compute
the contribution to $\left[\Psi_{55}\right]_0^{2\pi R}$ arising from the
zero mode only. This is done by integrating Eq.\ (\ref{rel.Sigma.55}),
which now simplifies to
\begin{equation}
\frac{3}{2} \Sigma'' - \frac{k^2}{a_0^2} \Sigma + \frac{3}{4} \Psi_{55}''
= M_*^{-3} \delta T^0_{\ 0} \ \delta(y) .
\end{equation}
Replacing $\Sigma$ by $\Sigma_{p=0}$,
and using the constraint $\int_0^{2 \pi R} \! dy \, \Psi_{55}=0$,
we get
\begin{equation}
(\Psi_{55})_{p=0} = \frac{2}{3} \frac{k^2}{a_0^2} 
 \left[(y - \pi R)^2  - \frac{1}{3} (\pi R)^2 \right]
\Sigma_{p=0}.
\end{equation}
By evaluating this relation at $y=0$ we find that the relation between $\phi_0$
and $\psi_0$ arising from the zero mode is
\begin{equation}
\psi_0 - \phi_0 = 
\frac{2 (k \pi R)^2}{9 a_0^2}
(\psi_0 + \phi_0) .
\label{rel.phi0.psi0}
\end{equation}
Let us first examine the limit $H R \ll k R / a_0 \ll 1$
in which we expect to recover
the four-dimensional results. Indeed, in this case we find from Eqs.\
(\ref{sigma.0}) and (\ref{rel.phi0.psi0}) that
\begin{equation}
\phi_0 = \psi_0 = \frac{1}{4 \pi R M_*^3} \
i \frac{a_0}{k} \dot{\bar{\varphi}} \delta \varphi
= \frac{1}{2 M_P^2} \ i \frac{a_0}{k} \ \dot{\bar{\varphi}} \delta \varphi ,
\end{equation}
which is exactly the four-dimensional result of Eq.\ (\ref{4d.sw}).  On the
other hand, in the limit $k R / a_0 \gg 1$ the contribution to $\phi_0$ and
$\psi_0$ arising from the zero mode is
\begin{equation}
- \phi_0 = \psi_0 = \frac{(k \pi R)^2}{ 9\ a_0^2}
\frac{1}{2 \pi R M_*^3} \ i \frac{a_0}{k} \dot{\bar{\varphi}} \delta \varphi .
\end{equation}
So, the right-hand side in the perturbed Klein-Gordon equation is of order
\begin{equation}
\left| \dot{\bar{\varphi}} ( \dot{\phi}_0 + 3 \dot{\psi}_0 ) \right|
= \left( 
\frac{\sqrt{2} k \pi R}{3 \ a_0}
\right)^2
\frac{\dot{\bar{\varphi}}^2}{2 \pi R M_*^3} 
\left|
\delta \varphi
\right| 
= \left(
\frac{2 k \pi R}{3 \ a_0}
\right)^2
\left| \dot{H} \delta \varphi
\right| \ll 
\left( 
\frac{2 \pi R H \ k}{3 a_0}
\right)^2
\left|
\delta \varphi
\right|,
\end{equation}
where we used the slow-roll inequality $\epsilon = -\dot{H}/H^2 \ll 1$.  If we
remember that $ \pi R H < 1$ and that the leading terms in the homogeneous
perturbed Klein-Gordon equation are of order $(k/a_0)^2 | \delta \varphi |$, we
see that even in this regime, the dynamics of the scalar field is unaffected by
that of the metric perturbations. This justifies the assumption that we made in
the long wavelength regime that the field dynamics is the same as in
four-dimensional physics (same vacuum normalization and same evolution).

\section{The consistency relation \label{consistencysection}}

Four-dimensional single-field models of inflation predict a consistency
relation \cite{reviewrocky} relating the amplitude of the scalar perturbations,
${\cal P}_S(k)$, the amplitude of the tensor perturbations, ${\cal P}_T(k)$,
and the tensor spectral index, $n_T\equiv d\ln {\cal P}_T(k)/d\ln k$.

Indeed, in four dimensions since ${\cal P}_T(k)\propto H^2(k)$, $n_T$ is given
by $n_T = d\ln H^2(k)/d\ln k = -2\epsilon$.  The four-dimensional consistency
relation is
\begin{equation}
\left. \frac{{\cal P}_T(k)}{{\cal P}_S(k)} \right|_{4D}= 16 \epsilon = -8 n_T .
\end{equation}
In the five-dimensional universe, however, both the amplitude and the tilt of
the tensor power spectrum receive corrections which are functions of $(RH)^2$.

If we parameterize the $R$-dependent corrections to the power spectrum of
tensor modes as 
\begin{equation}
{\cal P}_T(k) = \frac{2}{\pi^2} \left[\frac{H(k)}{M_{P}}\right]^2 
\frac{1}{1 - \alpha R^2 H^2(k)} ,
\end{equation}
(recall that our result was $\alpha=2\pi^2/3$) we may compute the spectral
index of the tensor modes to be
\begin{equation}
n_T = \frac{d\ln {\cal P}_T(k)}{d\ln H^2(k)}\frac{d\ln H^2(k)}{d\ln k} = 
\frac{1}{1 - \alpha R^2 H^2}
\frac{d\ln H^2(k)}{d\ln k} =-\frac{2\epsilon}{1 - \alpha R^2 H^2} ,
\label{t}
\end{equation}
where we have used the fact that the change of the Hubble parameter as a
function of scale, $d\ln H^2(k)/d\ln k$, is still given by $-2\epsilon$ since
the inflaton field is a brane field and the Hubble rate still satisfies the
four-dimensional equation $\dot{H} = -\epsilon H^2$. 

Using Eqs.\ (\ref{psh}), (\ref{cur}), and (\ref{t}), we find
\begin{equation}
\frac{{\cal P}_T(k)}{{\cal P}_S(k)} =
\frac{16\epsilon}{1 - \alpha R^2 H^2}=-8 n_T .
\end{equation}
This is a particularly surprising result: the consistency relation remains
unaltered at lowest order in the slow-roll parameters. A similar result has
been found in Ref. \cite{hl} for a set-up where the bulk on either side of the
brane corresponds to Anti-de Sitter {\it AdS} spaces with different
cosmological constants.  This degeneracy between the usual result in 4-D
one-field inflation and in extra-dimensional models will make it more difficult
to disentangle the various theoretical possibilities from observations. Our
results hold in the case in which only curvature perturbations are generated
during the inflationary phase. If isocurvature perturbations are produced, the
consistency relation in brane world scenarios is expected to differ from the
one obtained here as it happens in the four-dimensional case \cite{bmr}.

Of course this conclusion depends strongly on the particular form of the
corrections to the power-spectrum of the tensor perturbations, $(1 - \alpha R^2
H^2)^{-1}$, which holds in our five-dimensional example.  One can show that
this is actually the only possible functional dependence on $H$ such that the
consistency relation remains unaltered.

However, our result seems to be quite robust: even considering more than one
extra-dimension or even a Randall--Sundrum like scenario, we show in the next
section that the power spectrum of tensor perturbations always gets corrections
of the same form, as long as the radius is completely stabilized during
inflation.

Of course our calculation of the consistency relation is only to lowest-order
in slow-roll parameters.  In general, one expects the usual four-dimensional
corrections to the lowest-order result, corrections from the five-dimensional
background equations if $H$ is not constant, and corrections of order
$(H^2R^2)$ to the scalar perturbations as indicated in Eq.\ (\ref{eq:leading}).

\section{Generalization of the results}

We now would like to generalize some of our considerations to the case of more
than one extra dimension in the case in which the sizes of the extra dimensions
are all equal.  If we assume that the compactified geometry of the extra
dimensions is stabilized, we can take the background metric in the form of Eq.\
(\ref{backgroundmetric}) with $dy^2\equiv \delta_{\alpha \beta} dy^\alpha
dy^\beta$, $\alpha, \beta =1,\dots ,\delta$.  Equations (\ref{g00}) and
(\ref{g05}) for $\delta$ extra dimensions become
\begin{eqnarray}
G^0_{\ 0} & = & \frac{3}{a^2}\left( \frac{\dot{a}^2}{n^2}-\frac{1}{2}
\partial_\alpha \partial^\alpha a^2 \right) = M_*^{-(2+\delta)} \rho(t) \ 
\delta (y_1)\dots \delta (y_\delta),
\label{g00p} \\
G^0_{\ \alpha} & = & -\frac{3}{an} \partial_\alpha 
\left( \frac{\dot{a}}{n}\right) =0 .
\label{g05p}
\end{eqnarray}

Let us now make the simplifying assumption that because of rotational
symmetry in the extra dimensions the scale and lapse functions $a$ and $n$
depend only on the distance from the brane, $r\equiv (\sum_\alpha y_\alpha^2
)^{1/2}$, and on time, but not on the angular variables. Then, 
Eq.\ (\ref{g00p}) becomes
\begin{equation}
G^0_{\ 0} = -\frac{3}{2a^2}\left[ \frac{d^2a^2}{dr^2}+\frac{(\delta -1)}{r}
\frac{da^2}{dr}-2 K^2\right] = M_*^{-(2+\delta)}\rho(t)
\frac{1}{S_\delta r^{\delta -1}} \delta (r) ,
\label{g00pp}
\end{equation}
where $S_{\delta}$ is the surface of a unit-radius sphere in $\delta$
dimensions. The lapse function is given by $n=\dot{a}/K$, where $K$ is
independent of $r$, so we can choose $K= \dot{a}(\bar r )/n(\bar r)$, evaluated
at an arbitrary point $r=\bar r$. For $\delta >1$, the solution of Eq.\
(\ref{g00pp}) becomes singular at $r=0$, where the brane is located.

To overcome this difficulty, we define a brane with a finite thickness, and
impose our boundary conditions at $r=\epsilon$, keeping only the leading terms
in the limit $\epsilon \to 0$. We assume that appropriate density terms within
the brane smooth the singularity at the origin.  Next, we require
compactification conditions that, for simplicity, involve only the variable
$r$, and we impose that the values of the scale factor $a$ at $r=\epsilon$ and
at $r=2\pi R -\epsilon$ are equal.  Then, the solution of Eq.\ (\ref{g00pp}) is
\begin{equation}
a=a_\epsilon \sqrt{1-(\pi RH)^2 c(r)},
\end{equation}
with $c(r)$ given by
\begin{equation}
c(r) = \left\{
	\begin{array}{ll}
	 \Frac{2r}{\pi R} - \left( \Frac{r}{\pi R}\right)^2 
                           & \quad \textrm{for } \delta =1 \\ &  \\
	 \Frac{2 \ln r/\epsilon}{\ln 2\pi R/\epsilon} -\Frac{1}{2}
              \left( \Frac{r}{\pi R}\right)^2 & \quad \textrm{for } \delta =2 
                                                              \\ &  \\
	 \Frac{4}{\delta} \left[ 1-\left( \Frac{\epsilon}{r}\right)^{\delta 
               -2} \right] -\Frac{1}{\delta}\left( \Frac{r}{\pi R}\right)^2  &
               \quad \textrm{for } \delta \ge 3.
	\end{array}
\right.
\end{equation}
Here $a_\epsilon =a(r=\epsilon)$, and $H=\dot{a}_\epsilon / (n_\epsilon
a_\epsilon )$ is determined to be proportional to $\sqrt{\rho}$ by the jump
condition of $d a/dr$. In the case $\delta =1$ we can safely take the limit
$\epsilon \to 0$ and we reproduce Eq.\ (\ref{aandn}).

The correction factor for the tensor perturbations can be obtained following
the same argument used in Sec.\ \ref{tensorsection}. The effective
gravitational Planck mass during inflation is
\begin{equation}
M_P^2|_I =M_*^{\delta -2} \int d^\delta \!y  \ n^2= M_P^2\left[ 1- (\pi RH)^2
C\right],
\end{equation}
with
\begin{equation}
C\equiv \frac{\int_\epsilon^{2\pi R -\epsilon}  dr \ r^{\delta -1} c(r)}
{\int_\epsilon^{2\pi R -\epsilon}  dr \ r^{\delta -1} }.
\end{equation}
The coefficient $C$ depends on the compactification geometry. In the
five-dimensional case studied in Sec.\ \ref{tensorsection}, we found $C=2/3$,
while in the simplified case of $\delta$ extra dimensions compactified as
explained above, we obtain $C=8/[\delta (\delta +2)]$ (for $\delta \ge 2$).

In conclusion, our result that the tensor perturbations are enhanced by a
factor $\left[ 1- (\pi RH)^2 C\right]^{-1}$ is a generic consequence of our set
up with a stabilized geometry. The factor $C$ depends on the details of the
compactification and on the number of extra dimensions, but it is typically a
number of order unity. This form of corrections for the tensor perturbations
(together with the usual four-dimensional result for the scalar perturbations)
have the specific property of preserving the consistency relation, as discussed
in Sec.\ \ref{consistencysection}.

We can also extend our result to the case of non-factorizable
geometries \cite{rs}. Let us consider a 5-dimensional set up with two branes at
$y=0$ and $y=\pi R$, with vacuum densities equal in magnitude but opposite in
sign ($V_0=-V_\pi$), and a cosmological constant $\Lambda$ in the bulk. We
assume the usual relation between $V_0$ and $\Lambda$ to obtain the warped
Randall--Sundrum metric in vacuum, and we include a constant energy density
$\rho$ on the negative-tension brane. Imposing the condition that the
compactification radius is stabilized during the cosmological evolution, we
obtain that the scale and lapse functions are given by \cite{outstanding}
\begin{equation}
a=a_\pi \frac{n}{\Omega}, \qquad \Omega \equiv e^{-\pi KR},
\end{equation}
\begin{equation}
n^2=e^{-2K|y|}\left[ 1+\frac{(2\Omega^2 -1)H^2}{4\Omega^2K^2}\right]
+\left( \frac{e^{2K|y|}}{2}-1 \right) \frac{\Omega^2H^2}{2K^2},
\end{equation}
where the 5-dimensional coordinate $y$ is defined in the interval $(-\pi R,
\pi R)$. Here $K\equiv \Lambda /V_\pi$ is the inverse of the {\it AdS} 
radius, and $H \equiv {\dot a}_\pi / (n_\pi a_\pi)$ is the Hubble constant on
the visible brane located at $y=\pi R$, with $a_\pi =a(y=\pi R)$ and $n_\pi=
\Omega$. The Hubble constant is related to the energy density $\rho$ by the
equation
\begin{equation}
H^2=\frac{K\rho}{3M_*^3 (1-\Omega^2)}.
\end{equation}

We can now compute the Planck mass during inflation, which is given by
\begin{equation}
M_P^2|_I \equiv M_*^3 \int_{-\pi R}^{\pi R} dy \ n^2 =
\frac{M_*^3}{K} (1-\Omega^2) \left[ 1+\frac{H^2}{K^2} \left( 
\frac{3\Omega^2-1}{4\Omega^2}-\frac{\pi RK \Omega^2}{1-\Omega^2}\right)\right].
\end{equation}
Recalling that for the Randall--Sundrum model the Planck mass of the
4-dimensional effective theory is given by $M_P^2=(M_*^3/K) (1-\Omega^2)$,
we obtain in the limit of small warp factor $\Omega \ll 1$,
\begin{equation}
\left.M_P^2\right|_I =M_P^2 \left[ 1- \left(\frac{x_1H}{2m_1}\right)^2\right] .
\end{equation}
Here $m_1=x_1 K\Omega$ is the mass of the first graviton Kaluza--Klein mode (in
the limit $H=0$), and $x_1=3.8$ is the first root of the Bessel function $J_1$.
Therefore, the correction factor is quadratic in $H$ also in the case of
factorizable geometries, and the typical scale is determined by the
Kaluza--Klein mass gap.

We should note that in the case of the Randall--Sundrum model, the energy
separation between the Kaluza--Klein graviton mass and the fundamental scale at
which the theory becomes strongly interacting is often very small and this is a
limitation for the applicability of our result. Uncomputable quantum gravity
effects can become important and lead to comparable contributions.

\section{Conclusions}
In this paper we have initiated the investigation of the effects of
transdimensional physics on the spectrum of the cosmological density
perturbations generated during a period of primordial inflation taking place on
our visible three-brane. We have shown that the size of the transdimensional
effects are of order $(HR)^2$, where $H$ is the Hubble parameter during
inflation and $R$ is the typical size of the extra dimensions (or, more
precisely, the inverse of the Kaluza--Klein mass gap at zero temperature).  The
corrections appear in the power spectrum of the tensor modes. The coefficient
of the corrections depends upon the compactification geometry, the number of
extra dimensions, and if they are flat or warped.  As we have already stressed
in the Introduction, our treatment should be unaffected by (unknown) quantum
effects which might arise at distances below $M_*^{-1}$ as long as the size of
extra dimensions is larger than $M_*^{-1}$.

Our results may be generalized in different ways. First of all, our set up is
the simplest we could imagine: only one extra dimension and inflation taking
place on the brane.  One can envisage the possibility of putting the inflaton
field responsible for inflation in the bulk. In such a case we expect a
different form of corrections. In particular, the power spectrum of scalar
perturbations should be modified, thus possibly changing the consistency
relation.

One can also relax our working assumption of keeping the radii of extra
dimensions fixed. In this case there might be significant corrections to the
slope of the power spectra since having a dynamical radion field during
inflation amounts to change the Hubble rate during inflation.

In our paper we have also assumed that the energy density $\rho$ on the brane
is smaller than about $ M_{P}^2 / R^2$, or equivalently, that the Hubble radius
is larger than the radii of compactification. Deviations from the standard
four-dimensional Friedmann law are present in the opposite regime and large
deviations from the standard results for the power spectra of density
perturbations should appear.

Finally, we have assumed that deep in the ultraviolet regime, at distances much
smaller than the horizon length, the initial vacuum is the traditional
Bunch--Davies vacuum containing no initial particles in the spectrum.  This is
a reasonable assumption at physical momenta $k/a_0$ much larger than $R^{-1}$,
but still smaller than the fundamental scale $M_*$. Of course, for momenta
$k/a_0\gg M_*$, unknown quantum effects may take over and change drastically
the properties of the vacuum. This would lead to corrections scaling as powers
of $H/M_*$ as suggested by the analysis performed in the four-dimensional cases
\cite{tp}. Nevertheless, whenever $M_*R$ is sufficiently large, the computable
corrections discussed in this paper dominate over these unknown quantum
corrections.

\begin{acknowledgments}
A.R.\ would like to thank S.\ Matarrese, D.\ Lyth, and D.\ Wands for several
discussions and the Theory Group of CERN where part of this work was
done. J.L.\ acknowledges a visit to the University of Padua during which this
project was initiated.  The work of E.W.K.\ was supported in part by the
Department of Energy and by NASA (NAG5-10842).
\end{acknowledgments}



\appendix*

\section{Background and perturbed Einstein equations}

We give here the background and perturbed Einstein equations 
in the Gaussian normal gauge:
\begin{eqnarray}
G^0_{\ 0} & = & 
\frac{3}{n^2} \left( \frac{\dot{a}}{a} \right)^2 
- 3 \left[ \frac{a''}{a}
+ \left( \frac{a'}{a} \right)^2 \right] \nonumber \\
& & 
- \frac{6}{n^2} \frac{\dot{a}}{a}
\left( \frac{\dot{a}}{a} \phi + \dot{\psi} \right)
+ \frac{2}{a^2} \Delta \psi + \frac{2}{n^2 a^2} \frac{\dot{a}}{a} \Delta B
+ \frac{2}{n^2} \frac{\dot{a}}{a} \ \Delta \dot{E}
\nonumber \\
& & 
+ 3 \left( 4 \frac{a'}{a} \psi' + \psi'' \right)
- 4 \frac{a'}{a} \Delta E' - \Delta E''
\nonumber \\
& = & \delta(y) \left[ \left( \frac{\dot{\bar{\varphi}}^2}{2 n^2}
+ V \right) + \frac{\dot{\bar{\varphi}} 
\delta \dot{\varphi} - 
\dot{\bar{\varphi}}^2 \phi}{n^2} + V' \delta \varphi
\right] ,
\label{Einstein.00}
\\
G^i_{\ 0} & = & - a^{-2} \partial_i \left\{
2 \frac{\dot{a}}{a} \phi + 2 \dot{\psi}
+ \frac{2}{n^2} \left[ \frac{\ddot{a}}{a} 
- \left( \frac{\dot{a}}{a} \right)^2
- \frac{\dot{n}}{n} \frac{\dot{a}}{a} \right] B
\right.
\nonumber \\
& &
\left.
- \left( \frac{n''}{n} + \frac{a'}{a} \frac{n'}{n} \right) B 
+ \frac{1}{2} \left( \frac{a'}{a} - \frac{n'}{n} \right) B'
+ \frac{1}{2} B''
\right\}
\nonumber \\
& = & -a^{-2} \delta(y) \partial_i \left(
\dot{\bar{\varphi}} \delta \varphi - \frac{\dot{\bar{\varphi}}^2}{n^2} B
\right)
\\
\textrm{or  } G^0_{\ i} &=& n^{-2} \partial_i \left\{
2 \frac{\dot{a}}{a} \phi + 2 \dot{\psi}
\right.
\nonumber \\
& &
\left.
+ \left[ \frac{a'}{a} \frac{n'}{n} - 2 \left( \frac{a'}{a}\right)^2
- \frac{a''}{a} \right] B 
+ \frac{1}{2} \left( \frac{a'}{a} - \frac{n'}{n} \right) B'
+ \frac{1}{2} B''
\right\}
\nonumber \\
G^i_{\ j} & = & \left\{
  \frac{1}{n^2} \left[2 \frac{\ddot{a}}{a} 
+ \frac{\dot{a}}{a} \left(\frac{\dot{a}}{a} - 2 \frac{\dot{n}}{n}\right)\right]
-  2 \frac{a''}{a} - \frac{a'}{a}
\left( \frac{a'}{a} + 2 \frac{n'}{n} \right) - \frac{n''}{n} 
\right.
\nonumber \\
& &
 \left. - \frac{2}{n^2} \left[ 2 \frac{\ddot{a}}{a}
+ \left( \frac{\dot{a}}{a} \right)^2 - 2 \frac{\dot{a}}{a} \frac{\dot{n}}{n} 
\right] \phi - 2 \left( \frac{a'}{a} + \frac{n'}{n} \right) \phi'
- \phi'' - \frac{2}{n^2} \frac{\dot{a}}{a} \dot{\phi}
\right.
\nonumber \\
& &
\left.
+ 2 \left( \frac{n'}{n} + 3 \frac{a'}{a} \right) \psi' + 2 \psi'' 
- \frac{2}{n^2} \left( 3 \frac{\dot{a}}{a} - \frac{\dot{n}}{n} \right) 
\dot{\psi} - \frac{2}{n^2} \ddot{\psi}
\right\} \delta_{ij}
\nonumber \\
& & + a^{-2} (\delta_{ij} \Delta - \partial_i \partial_j) 
\left\{ -\phi + \psi 
+ \frac{1}{n^2} \left[ \left( \frac{\dot{a}}{a} 
- \frac{\dot{n}}{n} \right) B + \dot{B} \right]
+ \frac{a^2}{n^2} \left[ 
\left( 3 \frac{\dot{a}}{a} - \frac{\dot{n}}{n} \right) \dot{E} 
+ \ddot{E} \right] 
\right.
\nonumber \\
& &
\left.
- a^2 \left[ \left(3\frac{a'}{a} + \frac{n'}{n} \right) E'
+ E'' \right] \right\}
\nonumber \\
& = & \delta(y) \left( V -  \frac{\dot{\bar{\varphi}}^2}{2 n^2}
 +  \frac{\dot{\bar{\varphi}}^2 \phi -
\dot{\bar{\varphi}} \delta \dot{\varphi}}
{n^2} + V' \delta \varphi \right) \delta_{ij}
\\
G^0_{\ 5} & = & 
\frac{3}{n^2} \left[
\left( \frac{\dot{a}}{a} \frac{n'}{n} - \frac{\dot{a}'}{a} \right)
(1 - 2 \phi) + \frac{\dot{a}}{a} \phi' + \frac{\dot{a}}{a}
\psi' + \left( \frac{a'}{a} - \frac{n'}{n} \right) \dot{\psi}
+ \dot{\psi}' \right] \nonumber \\
& & + \frac{1}{n^2 a^2} \left( \frac{n'}{n} \Delta B 
- \frac{1}{2} \Delta B' \right)
\nonumber \\
& &
+ \frac{1}{n^2} \left[
- \Delta \dot{E}' + \left( \frac{n'}{n} - \frac{a'}{a} \right) \Delta \dot{E}
- \frac{\dot{a}}{a} \Delta E'
\right] = 0
\\
G^i_{\ 5} & = &  a^{-2} \partial_i\left\{\phi'-2\psi'
+\left({n'\over n}-{a'\over a}\right) \phi  
+ \frac{1}{n^2} \left[ \frac{a'}{a} \dot{B} -
\frac{1}{2} \dot{B}' + \frac{1}{2} \left(  \frac{\dot{n}}{n} -3 
\frac{\dot{a}}{a} \right) B'  
\right. \right.
\nonumber \\ 
& & \left.\left.
+\left( 2\frac{\dot{a}a'}{a^2}-2\frac{\dot{a}'}{a}-\frac{\dot{n}a'}{na}
+3\frac{\dot{a}n'}{an}\right)B
 \right] \right\} = 0 .
\label{Einstein.i5}
\end{eqnarray}
These equation simplify in the De Sitter background defined in Eq.\
(\ref{desitter}):
\begin{eqnarray}
n^2 \delta G^0_{\ 0} & = & 
- 6 H \left( H \phi + \dot{\psi} \right) + 2 \frac{\Delta}{a_0^2} \psi 
+ 2\frac{ H \Delta}{a_0^2n^2} B + 2 H \Delta \dot{E}
\nonumber \\
& & 
+ 3 \left( 4 n'n \psi' + n^2 \psi'' \right) - 4 n'n \Delta E' - n^2\Delta E''
\label{Einstein.DeSitter.00}
\\
- a^2 \delta G^i_{\ 0} &=& \partial_i \left( 2 H \phi + 2 \dot{\psi}
- \frac{1}{n^2} H^2 B + \frac{1}{2} B''
\right)
\\
n^2 \delta G^i_j &=& \left( - 6 H^2 \phi  - 4 n'n \phi' - n^2 \phi'' 
- 2 H \dot{\phi} \right.
\nonumber \\
& &
\left. + 8 n'n \psi' + 2 n^2 \psi'' - 6 H \dot{\psi} - 2 \ddot{\psi}
\right) \delta_{ij}
\nonumber \\
& & + (a_0n)^{-2} (\delta_{ij} \Delta - \partial_i \partial_j) 
\left[ -\phi + \psi + n^{-2} \left( H B + \dot{B} \right)
+ a_0^2 \left( 3 H \dot{E} + \ddot{E} \right) 
\right.
\nonumber \\
& &
\left.
- a_0^2 \left( 4 n'n E' + n^2 E'' \right) \right]
\\
n^2 \delta G^0_{\ 5} & = & 
3 \left( H \phi' + H \psi' + \dot{\psi}' \right) 
+ \frac{1}{a_0^2 n^2} \left( \frac{n'}{n} \Delta B - \frac{1}{2} \Delta B' 
\right) - \Delta \dot{E}'  - H \Delta E'
\\
- a^2 G^i_{\ 5} &=& \partial_i \left[ -\phi'+2\psi' \right.
\nonumber \\
& &
\left.
+ \frac{1}{n^2} \left(
\frac{1}{2} \dot{B}' + \frac{3}{2} H B' - \frac{n'}{n} \dot{B}
- 3 H \frac{n'}{n} B \right) \right] .
\label{Einstein.DeSitter.i5}
\end{eqnarray}


\begin{thebibliography}{99}
\bibitem{footnotea0}
Notice that $a_0(t)$ is the value of the scale factor at $y=0$, not its present
value.
\bibitem{eigenvalues}
This equation admits the solution $\omega =0$ for any value of $RH$. In the
limit of small $RH$, the solutions of Eq.\ (\ref{eigenvalue}) are $\omega_p =
p/R$, with $p$ ranging over the non-negative integers. Indeed, in the limit
$RH\to 0$, the space is not warped ($n \to 1$) and we recover the usual
equally-spaced Kaluza--Klein eigenvalues.  The solutions for $0<\pi RH<1$ are
plotted in Fig.\ 1.
\bibitem{effective}
Notice that if we defined the effective four-dimensional Planck mass during
inflation by integrating the background gravitational action over $y$, we would
obtain a different result, $\left.M_{P}^2\right|_{I}= M_{P}^2
\left(1 - \frac{1}{3} \pi^2 R^2 H^2\right)$. However, we are interested
here in the effective gravitational coupling of the zero Kaluza--Klein 
mode of the tensor perturbation, which is properly defined by
Eq.\ (\ref{efpl}).
\end{thebibliography}
\end{document}